How COVID-19 influences healthcare workers' happiness: Panel data analysis in Japan.

January 2021[1]

Eiji Yamamura* and Yoshiro Tsutsui

*Abstract*

Healthcare workers are more likely to be infected with the 2019 novel coronavirus (COVID-19) because of unavoidable contact with infected people. Although they are equipped to reduce the likelihood of infection, their distress has increased. This study examines how COVID-19 influences healthcare workers' happiness, compared to other workers. We constructed panel data via Internet surveys during the COVID-19 epidemic in Japan, from March to June 2020, by surveying the same respondents at different times. The survey period started before the state of emergency, and ended after deregulation. The key findings are as follows. (1) Overall, the happiness level of healthcare workers is lower than that of other workers. (2) The biggest disparity in happiness level, between healthcare workers and others, was observed after deregulation and not during the state of emergency. After deregulation, the difference was larger by 0.26 points, on an 11-point scale, than in the initial wave before the state of emergency.

[1] Corresponding author: Eiji Yamamura, Seinan Gakuin University, 6-2-92 Nishijin Sawaraku Fukuoka, 814-8511, Japan (e-mail:yamaei@seinan-gu.ac.jp). Coauthor: Yoshiro Tsutsui, Kyoto Bunkyo University, Japan (email: tsutsui@econ.osaka-u.ac.jp);. Acknowledgments: We would like to thank Editage [http://www.editage.com] for editing and reviewing this manuscript for English language.

# 1. Introduction

Since 2019, COVID-19 has spread rapidly from China to all over the world, and was officially recognized as a pandemic by the World Health Organization (WHO) on March 11, 2020. Responding to the ongoing COVID-19 crisis requires health policy analysts to conduct research and provide insights, to prepare for future crises [1]. COVID-19 adversely influences perception and physical condition [2,3]. It "acts as a stressor, causing health and economic anxiety, even in households that are not directly affected by the virus." [4]. Although policy makers need to address the issues raised by COVID-19, policies to mitigate its spread may result in a deterioration of mental health. On the one hand, in various countries (e.g., the United Kingdom, the United States, Italy, France), policymakers have implemented lockdowns to prevent the spread of the COVID-19 pandemic [5,6,7,8]. On the other hand, these stringent measures are considered to cause distress and increase domestic violence [9].

Similar to these countries, the Japanese government declared a state of emergency throughout Japan, although their policy was a less stringent measure than lockdown. The state of emergency promotes preventive behaviors, but exacerbates mental health [10]. In particular, "healthcare staff are at increased risk of moral injury and mental health problems when dealing with COVID-19" pandemic challenges [11]. Mental health is positively associated with labor productivity. [12]. Therefore, the mental health of healthcare workers must be maintained, to deliver quality and quantity of medical services. Researchers have highlighted the importance of guarding the mental health of healthcare workers [11,13]. Several studies have observed a higher prevalence of depression, stress and anxiety in medical healthcare workers, compared to non-medical healthcare workers, under the spread of COVID-19 [14, 15]. To the best of our knowledge, no studies have examined how the mental condition of healthcare workers changes during the spread of COVID-19. Therefore, this study uses data from Japan to explore how healthcare workers' subjective well-being changes, and to what extent this differs from other workers, in the course of the COVID-19 spread.

Individual-level panel data was collected via Internet surveys in Japan. The surveys were conducted five times during the three months, from March 13, to mid-June. During this period, the state of emergency was declared and subsequently deregulated. The questionnaire established respondents' happiness levels, as a proxy for subjective well-being. The contribution of this study is that it tracks changes in happiness levels, from before and after a period of stringent measures, such as a state of emergency. Key findings are that healthcare workers' happiness levels are lower than others. In particular, the largest disparity in happiness levels is observed after deregulation of the state of emergency. This implies that policies are needed to protect healthcare workers' mental health, especially after deregulation.

# 2. Setting and data

*2.1. Setting*

In Japan, the first person infected by COVID-19 was confirmed on January 16, 2020. Compared to other countries, such as Italy, France and Spain, the number of infected people increased over time, but at a much slower rate. To investigate the impact of COVID-19 on people's happiness levels, we conducted Internet surveys to gather data.

[Insert Figure 1 here]

The Internet survey included questionnaires, sent to selected Japanese citizens, aged 16 to 79, throughout Japan. INTAGE, which is a company with abundant experience in academic research, was commissioned for conducting the Internet survey. In the survey, we ordered to use the sampling method to gather a representative sample of the Japanese population regarding gender, age, and prefecture of residence. Figure 1 illustrates the survey time points and the state of emergency period, accompanied by the number of daily infections. To construct the panel dataset, the survey was conducted five times, to track the same individuals for almost three months, from March to June. As shown in Figure 1, there were only 40 daily infections on March 13, when the first wave of the survey was conducted. Accordingly, when the survey started, COVID-19 had little impact on daily life in Japan. There was a surge in the number of daily infections before the declaration of the state of emergency. After entering the state of emergency, the number of daily infections fell to less than 100, similar to the level of the first wave.

In the first wave, 4,359 observations were gathered and the response rate was 54.7%. The second, third and fourth waves were conducted on March 27, April 10, May 8, and June 12, respectively. The response rates reached 80.2% (second wave), 92.2% (third wave), 91.9% (fourth wave), and 89.4% (fifth wave). The total number of observations was 19,740. The sample was further limited to those with employment. Thus, the sample size was reduced to 10,901. Respondents were asked about happiness level, occupation, education, household income, marital status, and gender. In Table A 1 presented in Online Appendix, the sample is restricted to workers, and its definition, mean values, and standard deviations are given. This study focuses on the different situations people were working in, when investigating the effects of COVID-19. The mean value of *health care workers,* a dummy of healthcare workers, is 0.07, indicating that 7% of respondents worked in healthcare. Unlike existing studies that identify medical healthcare workers [13,14], this study classifies *health care workers* not only as medical workers, but also as non-medical workers, such as nursing caregivers. However, both medical and non-medical workers may have physical contact with a various people in the workplace, subsequently increasing the risk of contracting COVID-19.

Figure 1 shows that the Japanese government declared a state of emergency in April, between the second and third wave. The declaration mandated people to avoid going out unnecessarily, and for the closing of various public places including schools, museums, theaters, and bars, among others. Immediately after the declaration, the third wave of the survey was conducted. On 25 May, 2020, the state of emergency was deregulated as daily infections were reduced to between 20 and 30. Therefore, the advantage of the data is that it covers the period when the situation changed significantly.

[Insert Figure 2 here]

Figure 2 compares the mean values of happiness in each wave, between healthcare workers and other workers. Generally, the happiness level was highest in the first wave, then decreased in the second and third waves. Subsequently, the happiness level increased in fourth and fifth waves. In all waves, the happiness level of healthcare workers is lower than others. The disparity is smallest in the first wave, increases in the second and third waves, and reduces from the third to fourth waves. However, it increases considerably from the fourth to the fifth wave. Consequently, the disparity is largest in the fifth wave.

This implies that general happiness levels declined as COVID-19 spread, until the declaration of the state of emergency. Considered together, Figures 1 and 2 indicate that, in particular, healthcare worker's happiness declines as daily infections increase. Interestingly, although the state of emergency was deregulated, healthcare workers' happiness level remains unchanged, whereas it increases for others. In the following section, we use a regression method to scrutinize happiness levels, after controlling for various factors.

## 3. Method

As the baseline model, the estimated function is as follows:

$$Happiness_{it} = \alpha_0 + \alpha_1\ Health\ Care\ Worker_i + \alpha_2\ Schooling\ years_i + \alpha_3\ Income_i + \alpha_4\ COVID\text{-}19_{it} + \alpha_5\ Marry_i + \alpha_6\ Female_i + \alpha_7\ Ages_{it} + k_t + u_{it}.$$

In this formula, $Happiness_{it}$ represents the dependent variable of individual $i$, wave $t$. An ordinary least square (OLS) model is used. Using pooled samples, the effect of data collection was controlled by including dummies for the second ($Wave2$), third ($Wave3$) fourth ($Wave4$), and fifth waves ($Wave5$), while the reference group was the first wave. In addition, separate estimations were made using subsamples of each wave. The key variable is the *health care worker*. The sign of the coefficient is expected to be negative, because healthcare workers are considered to be more distressed than other workers. The absolute values of the coefficient are thought to change with the subsamples, reflecting different situations. In addition, various control variables are included in demographic and economic factors, which are usually included in the analysis of subjective well-being [16,17]. The regression parameters are denoted as $\alpha$. The number of COVID-19 infections dramatically increased in residential areas during the study period. Therefore, this was included as a control variable. The error term is denoted as $u$.

For closer examination, we controlled for individual fixed effects that are time-invariant respondent features [18]. The estimated function is as follows:

$$Happiness_{it} = \alpha_1\ Wave2_t \times Health\ Care\ Worker_i + \alpha_2\ Wave3_t \times Health\ Care\ Worker_i + \alpha_3\ Wave4_t \times Health\ Care\ Worker_i + \alpha_4\ Wave5_t \times Health\ Care\ Worker_i + \alpha_5\ Wave2_t + \alpha_6\ Wave3_t + \alpha_7\ Wave4_t + \alpha_8\ Wave5_t + \alpha_4\ Infected\ COVID19_{it} + m_i + u_{it}.$$

In this specification, $m_i$ is the individual's characteristics that remain unchanged over time; that is, $m_i$ is the date of birth and gender of the respondent. As the data used in this study included respondent characteristics from the first wave only, the following factors are assumed to have remained unchanged: annual household income, education level and marital status. This assumption is plausible because of the three-month survey period. The fixed-effects method can be controlled [17,18].

The key variables are cross-terms between *health care workers* and wave dummies, such as *Wave2 $_t$ × Health Care Worker, Wave3 $_t$ × Health Care Worker, Wave4 $_t$ × Health Care Worker, Wave5 $_t$ × Health Care Worker*. These variables reflect the manner and the extent of the difference, in the first wave, between the effects on *health care workers* and other workers. For example, if the coefficient of *Wave3 $_t$ × Health Care Worker* shows a negative sign and statistical significance, the difference in the third-wave happiness level, between healthcare workers and other workers, is less than the difference in the first wave. As explained in Section 2, the state of emergency was declared between the second and third waves. Therefore, the results indicate that healthcare workers became unhappier during the state of emergency, than before the state of emergency.

# 4. Results

The results presented in column (1) of Table 1 are based on the complete sample, from the first wave to the fifth wave. The results in columns (2) to (6) are based on the subsamples of each wave. Therefore, the wave dummy results are only reported in column (1). In addition to the variables shown in Table 1, dummies for respondents' residential prefectures and city sizes are included as independent variables.

[Insert Table 1 here]

Interpreting the results requires focus on the results of the key variable, *Health Care Worker*. In all columns, *Health Care Worker* shows negative signs. The complete sample result is statistically significant at the 5% level. This indicates that healthcare workers are unhappier than other workers, which is consistent with previous studies [13, 14]. For the subsample results, statistical significance is observed in columns (4) and (6). In particular, column (6) is significant at the 1% level. Moreover, in columns (4) and (6), the absolute values of the *Health Care Worker* coefficient are 0.28 and 0.54, respectively. This indicates that, in the first wave, the happiness level of healthcare workers is 0.54 points lower, on the 11-point scale, than other workers; whereas, in the third wave, the happiness level is only 0.28 points lower. Surprisingly, healthcare workers became unhappier after the state of emergency was deregulated, not immediately after the state of emergency was declared.

*COVID-19* shows a significant negative sign in five of the six results. For convenience of interpretation, the coefficients are multiplied by 1000. The absolute value of the coefficient is 4.68 in column (4), which is clearly the largest. This means that, with an increase of 1000 COVID-19 patients, there is a 4.68 decrease in happiness on the 11-point scale. Considering Table 1 and Figure 1 together, shows that happiness level was significantly influenced by information about the prevalence of COVID-19, during peak daily infections. On the other hand, the absolute value of the coefficient is 1.10 in column (6), which is the lowest subsample result. That is, daily infections were clearly reduced, after the state of emergency deregulation. Interestingly, column (1) indicates that the coefficient of *COVID-19* was not statistically significant in the complete sample results. Unlike the subsample estimations, the wave dummies were included in the complete sample estimation. *Wave2, Wave3, Wave4* and *Wave5*, all show negative signs and statistical significance at the 1% level, in the complete sample result. The absolute values of the coefficients of *Wave3* and *Wave4*, were approximately 0.37 and 0.36, more than double the coefficients of *Wave2* and *Wave5*. This implies that, during the state of emergency, people were approximately 0.37 points unhappier than at the start of the survey. Compared to when the survey started, workers were also unhappier directly before the state emergency declaration, and after its deregulation. However, the decline in happiness level, under the state of emergency, was far greater than directly before, and after, the state of emergency. Overall, for the complete sample results, the reason for the insignificant sign of *COVID-19* is the effect of the spread of COVID-19, captured by the wave dummies.

For other control variables, the results are in good agreement with those of existing studies investigating happiness levels [16, 17].

[Insert Table 2 here]

Table 2 shows the results of the fixed effects estimation. Cross terms show negative signs and statistical significance, except for *Wave4 $_t$ × Health Care Worker.* As shown in Table 1, the medical workers' happiness level is lower than that of other workers. Tables 1 and 2, considered together, indicate that the disparity in happiness level between healthcare workers and other workers increases with time. The absolute values of the cross-term coefficients were approximately 0.20, for *Wave2 × Health Care Worker* and *Wave3 × Health Care Worker.* This implies that the disparity in the second and third waves is approximately 0.20 points larger than the first wave. The value increased to 0.26 points for *Wave5 × Health Care Worker,* although there is no significant disparity for *Wave4 × Health Care Worker.* This is consistent with Figure 2 and Table 1.

## 5. Discussion

The findings in this study suggest that an increased number of infections poses a higher risk to healthcare workers than other workers, as healthcare workers are required to make physical contact with patients. Consequently, the happiness level of health of healthcare workers is lower than that of other workers. In the subsequent phase, people tended to adapt to the state of emergency and consequently, over time, developed preventive behaviors. This reduces the risk faced by healthcare workers.

However, the general population is more likely to go out once the state of emergency is deregulated. At the time of deregulation, on May 25, the total number of deaths caused by COVID-19 was only 830, far less than other developed countries, such as 99,267 (USA) and 6,872 (UK). Government policy on COVID-19 has been criticized for its rigor in producing an economic recession. After deregulation, the government directed the promotion of economic activity. For example, the Japanese central government implemented a campaign to promote domestic tourism [19], and the general population gradually returned to daily life, as before the spread of COVID-19. To subsidizes up to 50% of the eligible domestic travel costs, the government of Japan promoted the "Go To Travel" campaign which started July 22, prior to the four-day weekend. Consequently, the number of infections again increased significantly. After deregulation, healthcare workers predicted that COVID-19 would re-emerge. In summary, the disparity between healthcare workers and others' perspectives, regarding COVID-19, significantly increased after deregulation, and inevitably increased the disparity in happiness levels between them.

Good supervisor communication can improve employee mental health, thereby increasing productivity in the workplace [12]. Accordingly, healthcare managers must take active measures to protect the mental health of employees [11]. Policies should be adopted to strengthen altruism and prosocial behavior that can increase the chances of eliciting social support from others [20].

## 6. Conclusion

Patients under the COVID-19 pandemic should be provided with adequate quantity and quality of medical services. Healthcare workers play a critical role in the response to COVID-19. For this, we need to pay great attention to their mental health.

This study examines how COVID-19 influences healthcare workers' happiness levels, compared to other workers. The advantage of the data is that the panel structure can be used to track the same person, to explore variations in happiness levels, during rapid changes in circumstances. First, overall, we found that healthcare workers' happiness level is lower than others. Second, it is interesting to note that the difference between healthcare workers' and other workers' happiness levels was not greatest during the state of emergency, but actually after deregulation.

Our findings provide compelling evidence that, when society's crisis atmosphere for the pandemic was eased, healthcare workers were distressed and unhappy. Therefore, in order to provide quality medical services and save lives, health policies are required to raise awareness of the importance of alleviating the distress of healthcare workers. The government should adopt policies that allow society to share the sense of danger with healthcare workers. To bridge the gap between healthcare workers and other workers, policies should strengthen collective action to adopt to the new normal under the COVID-19 pandemic, rather than returning to normal life before COVID-19. Moreover, policies must be formulated to train healthcare managers to communicate with staff, to improve mental health in the workplace, especially in such situations as a pandemic [11].

Owing to data limitations, it was not possible to further classify healthcare workers into medical and non-medical workers. Further research should examine how medical workers' happiness level differs from that of non-medical workers, during the spread of COVID-19. In other countries, where COVID-19 became widespread, more stringent measures were adopted than in Japan. Further studies are needed to investigate whether the findings of this study were also observed in other countries under COVID-19. In this study, happiness level is considered to reflect mental health. The relationship between happiness and mental health, under COVID-19, should be examined. These are issues to be addressed in future research.

# References

[1] Bal R, de Graaff B, van de Bovenkamp H, Wallenburg I. Practicing corona - Towards a research agenda of health policies. Health Policy 2020;124:671–3. https://doi.org/10.1016/j.healthpol.2020.05.010

[2] Fetzer T, Witte M, Hensel L, Jachimowicz J, Haushofer J, Ivchenko A et al. Global behaviors and perceptions in the COVID-19 pandemic. PsyArXiv. 2020 April 16. https://doi:10.31234/osf.io/3kfmh

[3] Layard R, Clark A, De Neve JE, Krekel C, Fancourt D, Hey N, O'Donnell G. When to release the lockdown? A wellbeing framework for analysing costs and benefits. IZA Discussion Paper 2020;13186.

[4] Sabat I, Neuman-Böhme S, Varghese NE, Barros PP, Brouwer W, van Exel J, Schreyögg J, Stargardt T. United but divided: Policy responses and people's perceptions in the EU during the COVID-19 outbreak. Health Policy 2020. https://doi.org/10.1016/j.healthpol.2020.06.009

[5] Anderson R, Heesterbeek H, Klinkenberg D, Hollingsworth T. How will country-based mitigation measures influence the course of the COVID-19 epidemic? The Lancet 2020;395:931–4.

[6] Fang H, Wang L, Yang Y. Human mobility restrictions and the spread of the novel coronavirus (2019-nCoV) in China. NBER Working Paper 2020;26906. https://doi:10.3386/w26906

[7] Tian, H, Liu Y, Li Y, Wu C, Chen B, Kraemer, M et al. An investigation of transmission control measures during the first 50 days of the COVID-19 epidemic in China. Science 2020: https://doi:10.1126/ science.abb6105

[8] Viner R, Russell S, Croker H., Packer J, Ward J, Stansfield C et al. School closure and management practices during coronavirus outbreaks including COVID-19: A rapid systematic review. The Lancet Child and Adolescent Health 2020;4:397–404.

[9] World Health Organization (WHO). COVID-19 and violence against women: what the health sector/system can do. 2020. https://apps.who.int/iris/bitstream/handle/10665/331699/WHO-SRH-20.04-eng.pdf

[10] Yamamura E, Tsutsui, Y. Impact of the state of emergency declaration for Covid-19 on preventive behaviours and mental conditions in Japan: Difference-in-differences analysis using panel data. COVID Economics: Vetted and Real-Time Papers 2020;23:303–324.

[11] Greenberg N, Docherty M, Gnanapragasam S, Wessely S. Managing mental health challenges faced by healthcare workers during covid-19 pandemic. BMJ 2020;368:m1211. http://dx.doi.org/10.1136/bmj.m1211

[12] Kuroda S, Yamamoto I. Good boss, bad boss, workers' mental health and productivity: Evidence from Japan. Japan and the World Economy 2018: 48(C), 106–18. https://doi.org/10.1016/j.japwor.2018.08.002

[13] Chen Q, Liang M, Li Y, Guo J, Fei D, Wang L, He L, Sheng C, Cai Y, Li X, Wang J, Zhang Z. Mental health care for medical staff in China during the COVID-19 outbreak. The Lancet Psychiatry 2020;7(4),e15–6. https://doi.org/10.1016/S2215-0366(20)30078-X

[14] Zhang W, Wang K, Yin L, Zhao WF, Xue Q, Peng M et al. Mental health and psychosocial problems of medical health workers during the COVID-19 epidemic in China. Psychotherapy and Psychosomatics 2020;89:242–50. https://doi.org/10.1159/000507639

[15] Benjamin YQ, Tan MD, Nicholas WS, Chew MD, Grace KH, Lee MD et al. Psychological impact of the COVID-19 pandemic on health care workers in Singapore. Annals of Internal Medicine 2020. https://doi.org/10.7326/M20-1083

[16] Clark A, Oswald A. Satisfaction and comparison income. J Public Econ 1996;61(3):359–81.

[17] Yamamura E, Tsutsui Y, Yamane C, Yamane S, Powdthavee N. Trust and happiness: Comparative study before and after the Great East Japan Earthquake. Soc Indic Res 2015;123(3):919–935.

[18] Wooldridge, M.W. Econometric analysis of cross section and panel data. MIT Press; 2010.

[19] Japan Times. Let's discuss the Go To Travel campaign 2020. https://www.japantimes.co.jp/life/2020/07/21/language/english-lets-go-to-travel-campaign/.

[20] Shimazu A, Nakata A, Nagata T, Arakawa Y, Kuroda S, Inamizu N, Yamamoto I. Psychosocial impact of COVID-19 for general workers. J Occup Health 2020;62:e12132. https://doi.org/10.1002/1348-9585.12132

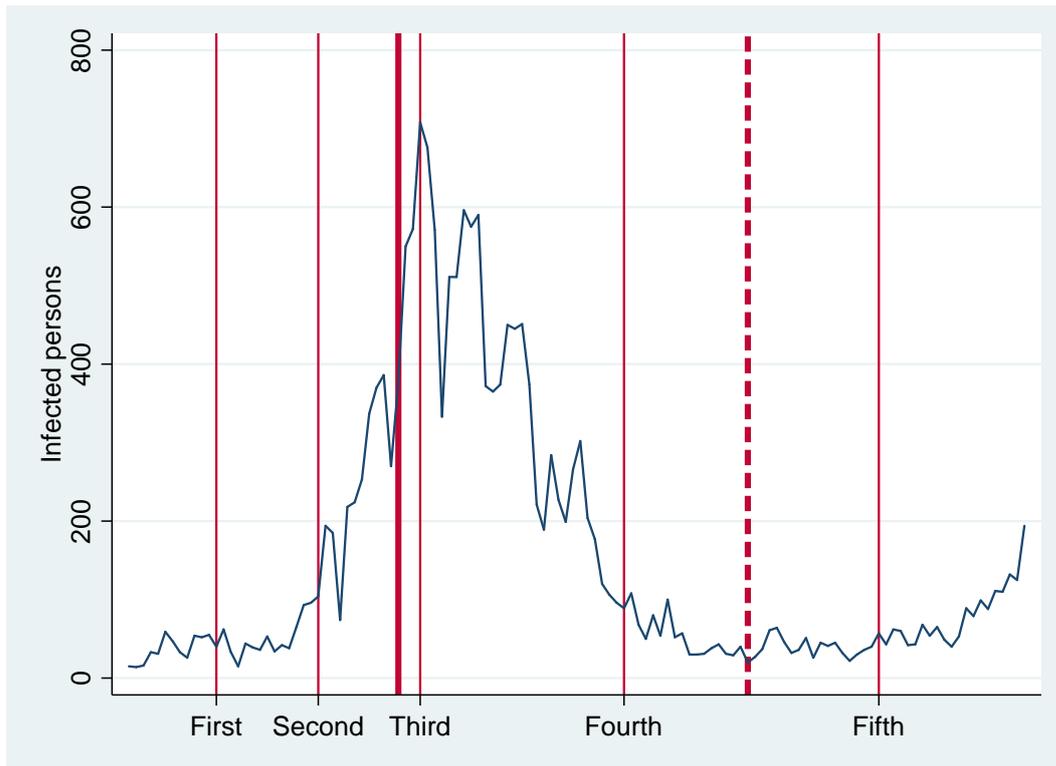

**Fig 1.** Changes in daily COVID-19 infections in Japan, March 1 to July 2.

Note: The first, second, third, fourth, and fifth waves were conducted on March 10, March 27, April 10, May 8, and June 12, respectively. The thin lines show the survey dates. The thick solid line indicates the date when the Japanese government declared the state of emergency (April 7). The thick dashed line indicates the date the state of emergency was deregulated (May 25).

Source: Daily COVID 19 infections are sourced from official website of the "Ministry of Health, Labour and Welfare." https://www.mhlw.go.jp/stf/covid-19/open-data.html. (Accessed July 4, 2020).

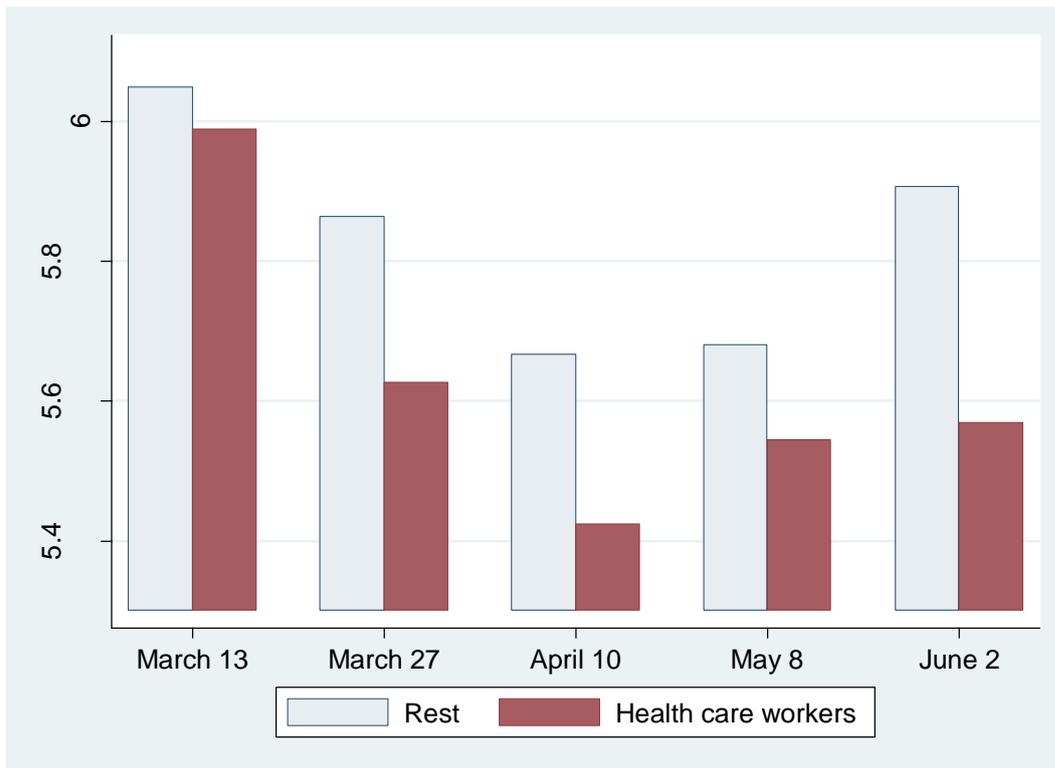

**Fig 2.** Mean values of Happiness in each wave.

Table 1. Baseline results. Dependent variables: Happiness level (OLS model)

|  | (1) Wave 1-5 | (2) Wave 1 | (3) Wave 2 | (4) Wave 3 | (5) Wave 4 | (6) Wave 5 |
|---|---|---|---|---|---|---|
| Healthcare Worker | −0.27** (0.13) | −0.15 (0.13) | −0.20 (0.15) | −0.28* (0.16) | −0.15 (0.17) | −0.54*** (0.17) |
| Schooling Years | 0.03* (0.02) | 0.03* (0.02) | 0.04** (0.02) | 0.04** (0.02) | 0.006 (0.02) | 0.03* (0.02) |
| Income | 0.74*** (0.10) | 0.78*** (0.20) | 0.74*** (0.14) | 0.68*** (0.14) | 0.72*** (0.14) | 0.78*** (0.13) |
| COVID-19 | 0.01 (0.01) | −1.95*** (0.70) | −1.36** (0.66) | −4.68*** (0.39) | −1.20** (0.10) | −1.10*** (0.07) |
| Married | 0.79*** (0.09) | 0.84*** (0.11) | 0.70*** (0.13) | 0.77*** (0.09) | 0.79*** (0.10) | 0.86*** (0.10) |
| Female | 0.56*** (0.09) | 0.59*** (0.11) | 0.59*** (0.10) | 0.46*** (0.11) | 0.54*** (0.11) | 0.61*** (0.09) |
| Ages | 0.01*** (0.002) | 0.01*** (0.003) | 0.01*** (0.003) | 0.02*** (0.003) | 0.01*** (0.003) | 0.01*** (0.003) |
| Wave1 | < default > |  |  |  |  |  |
| Wave2 | −0.17*** (0.03) |  |  |  |  |  |
| Wave3 | −0.37*** (0.04) |  |  |  |  |  |
| Wave4 | −0.36*** (0.05) |  |  |  |  |  |
| Wave5 | −0.13*** (0.04) |  |  |  |  |  |
| R-Square | 0.11 | 0.12 | 0.12 | 0.12 | 0.12 | 0.12 |
| Obs. | 10,689 | 2,344 | 1861 | 2,188 | 2,173 | 2,123 |

Note: Numbers within parentheses are robust standard errors clustered by residential prefectures. Dummies for respondents' residential prefectures and city size were included, although the results were not shown. ***, **, and * indicate statistical significance at the 1%, 5%, and 10 % levels, respectively. For convenience of interpretation, the coefficients of *income* and *COVID-19* were multiplied by 1000.

**Table 2.** Dependent variables: Happiness (Fixed effects model).

| | (1) |
|---|---|
| Wave2 ×Health Care Worker | −0.21** (0.09) |
| Wave3 ×Health Care Worker | −0.20** (0.10) |
| Wave4 ×Health Care Worker | −0.10 (0.11) |
| Wave5 ×Health Care Worker | −0.26** (0.102 |
| Wave2 | −0.14*** (0.04) |
| Wave3 | −0.34*** (0.04) |
| Wave4 | −0.32*** (0.04) |
| Wave5 | −0.07 (0.04) |
| Infected COVID_19 | −0.001 (0.02) |
| Within R-Square | 0.02 |
| Groups | 2,434 |
| Obs. | 10,901 |

Note: Numbers within parentheses indicate robust standard errors clustered by individuals. ***, ***, * indicate statistical significance at a level of 1%, 5%, and 10%, respectively.

Online Appendix

Table A 1. Definitions of key variables and their basic statistics

|  | Definition | Mean | Standard deviation |
|---|---|---|---|
| *Happiness* | "To what degree are you currently feeling happiness?" Please answer on a scale from 1 (very unhappy) to 11 (very happy) | 5.82 | 2.19 |
| *Healthcare Worker* | Equal to one if the respondent is a healthcare worker, otherwise equal to zero | 0.07 | 0.24 |
| *Schooling Years* | Respondent's schooling years | 14.1 | 2.16 |
| *Income* | Respondent's annual household income (million yen) | 5.46 | 3.78 |
| *COVID_19* | Total number of COVID-19 infections in respondent's residential prefecture | 527 | 1,105 |
| *Marry* | Equal to one if the respondent is married, otherwise equal to zero | 0.62 | 0.48 |
| *Female* | Equal to one if the respondent is female, otherwise equal to zero | 0.50 | 0.50 |
| *Age* | Respondent's age | 49.5 | 17.0 |
| *Wave 2* | Equal to one if survey is second wave, otherwise equal to zero | 0.20 | 0.20 |
| *Wave 3* | Equal to one if survey is third wave, otherwise equal to zero | 0.20 | 0.20 |
| *Wave 4* | Equal to one if survey is fourth wave, otherwise equal to zero | 0.20 | 0.20 |
| *Wave 5* | Equal to one if survey is third wave, otherwise equal to zero | 0.20 | 0.20 |

Note: The sample is limited to workers.